# Atmospheric Electricity at the Ice Giants

K. L. Aplin (1), G. Fischer (2), T. A. Nordheim (3), A. Konovalenko (4), V. Zakharenko (4), and P. Zarka (5)

1. Department of Aerospace Engineering, University of Bristol, UK
   *corresponding author* karen.aplin@bristol.ac.uk, +44 117 4283371
2. Space Research Institute, Austrian Academy of Sciences, Austria
3. Jet Propulsion Laboratory, California Institute of Technology, Pasadena, CA, USA
4. Institute of Radio Astronomy, National Academy of Sciences of Ukraine, Ukraine
5. Observatoire de Paris, Centre National de la Recherche Scientifique, PSL, France

**Abstract**
Lightning was detected by Voyager 2 at Uranus and Neptune, and weaker electrical processes also occur throughout planetary atmospheres from galactic cosmic ray (GCR) ionisation. Lightning is an indicator of convection, whereas electrical processes away from storms modulate cloud formation and chemistry, particularly if there is little insolation to drive other mechanisms. The ice giants appear to be unique in the Solar System in that they are distant enough from the Sun for GCR-related mechanisms to be significant for clouds and climate, yet also convective enough for lightning to occur. This paper reviews observations (both from Voyager 2 and Earth), data analysis and modelling, and considers options for future missions. Radio, energetic particle and magnetic instruments are recommended for future orbiters, and Huygens-like atmospheric electricity sensors for in situ observations. Uranian lightning is also expected to be detectable from terrestrial radio telescopes.

1. Introduction

The fundamental force of electricity is common in planetary atmospheres, with cosmic rays a ubiquitous source of ionisation (e.g. Aplin, 2006), and lightning detected in 4 ± 1 of 7 Solar System planets (Harrison et al, 2008). The ice giant planets Uranus and Neptune with their deep, cloudy atmospheres are both thought to have lightning, based on positive detections by the only spacecraft to have visited them, Voyager 2, in the 1980s (Aplin and Fischer, 2017). Galactic cosmic rays (GCRs) can penetrate and affect deep planetary atmospheric layers. These highly energetic particles originate from beyond our Solar System and initiate extensive cascades of secondary particles which deposit energy into ionisation along their paths. This ionisation acts as a source of charge, which enables atmospheric electrical processes.

Heating from lightning triggers chemical reactions that were demonstrated to produce amino acids in an Earth-like atmosphere (e.g. McCollom, 2013), indicating that lightning could be implicated in the origins of life. Though it is unlikely that this particular process could occur in the hydrogen and helium ice giant atmospheres, lightning provides energy for chemical reactions which could be significant in the outer solar system where there is little insolation. As the only known underlying cause of lightning is atmospheric convection, unambiguous detection imply convection and can therefore provide insight into atmospheric dynamics. The detection of Saturn lightning from a terrestrial radio telescope (Konovalenko et al, 2013) has encouraged the possibility that Uranian lightning may be detectable from Earth, particularly as giant storms on Uranus are observed from Earth using both ground-based facilities and the Hubble Space Telescope (de Pater et al, 2015).

Away from thunderstorms, the ions, electrons and other charged particles created by GCR make the air slightly electrically conductive, which can assist cloud formation, microphysics, and atmospheric chemistry. At planets like Earth which are relatively close to the Sun, insolation-driven processes dominate weather and climate. Uranus and Neptune, located at 20 and 30 AU respectively, receive a solar flux that is two to three orders of magnitude lower than on Earth, but a similar GCR flux, implying a proportionally greater role for electrical processes in their atmospheres. This is supported by spacecraft and ground-based data and modelling, indicating that charged aerosol particles and electrical effects play significant roles in ice giant atmospheres (e.g. Aplin and Harrison, 2016, 2017).

In this paper we will discuss the processes outlined above in more detail to provide an up-to-date review of the role and status of atmospheric electricity at the ice giant planets. We consider lightning generation and observations, both space and ground-based in section 2, then non-thunderstorm electricity in section 3. In section 4 we synthesise the discussion and use this to make recommendations for future measurements and instrumentation, both Earth and space based.

**2. Lightning**

*2.1 Brief overview of lightning detection technologies for ice giants*

The easiest way to detect lightning at the ice giants is by observing their electromagnetic emissions with antennas. The ionized lightning channel itself acts as an antenna and radiates electromagnetic waves over a broad frequency range from a few Hz up to several GHz (Rakov and Uman, 2003). At the lowest frequencies of a few Hz, lightning radio emissions can produce standing waves called Schumann resonances in the ionospheric cavity of a planet. However, their intensity is low and classical theory indicates that a sensitive in situ detector is needed, as the frequency is normally considered too low for the waves to escape the "ionospheric cutoff" (there is some evidence for a "leaky" ionosphere, permitting remote sensing of terrestrial Schumann resonances (Simoes et al, 2011)). Schumann resonances were detected on Titan but are attributed to a non-atmospheric electricity cause (Béghin et al, 2007), so this type of data needs to be carefully interpreted. A probe delivered to an ice giant atmosphere should thus have a lightning detector in the very low frequency range (VLF, 3-30 kHz), because lightning radio emissions are much stronger at these frequencies and can propagate over several thousands of kilometres within the ionospheric cavity. For example, the Galileo probe detected VLF bursts attributed to lightning with its lightning and radio emission detector (LRD) during its descent into the Jovian atmosphere. The LRD used a ferrite-core magnetic radio frequency antenna from 100 Hz to 100 kHz (Lanzerotti et al., 1992; Rinnert et al., 1998). VLF signals from lightning can also be detected from outside the planet's ionosphere in the form of whistlers, which are electromagnetic waves guided along magnetic field lines. Whistlers detected by the Voyager 2 plasma wave instrument around 6-12 kHz are the most important indication for lightning on Neptune (Gurnett et al., 1990), and we will discuss this observation in more detail in the next subsection.

The radio emissions from lightning called "sferics" can also be detected in the high frequency (HF) range (3-30 MHz). Such HF sferics, whose frequency is above the ionospheric cutoff, can pass directly through the ionosphere and freely propagate to orbiting spacecraft. Prominent HF sferics were detected at Saturn and (incorrectly) named "Saturn Electrostatic Discharges" (SED, Warwick et al., 1981), and at Uranus, where they were analogously called "Uranian Electrostatic Discharges" (UED, Zarka and Pedersen, 1986). SED were detected by the radio instruments on Voyagers 1 and 2 (Zarka and Pedersen, 1983), by Cassini (Fischer et al., 2008), and by ground-based telescopes (Konovalenko et al., 2013). UED were only detected by Voyager 2 (Zarka and Pedersen, 1986); SED and UED are compared in the next subsection. The spacecraft used electric monopole or dipole antennas and corresponding receivers for radio wave reception (Warwick et al., 1977; Gurnett et al., 2004). In the HF range the receivers swept through the frequencies with step increments of a few hundred kHz and dwelled at each frequency for several tens of milliseconds.

Another interesting detection was made recently with the Juno Microwave Radiometer (MWR) which detected impulses attributed to Jovian lightning at frequencies of 600 and 1200 MHz with a receiver bandwidth of 18 MHz (Brown et al., 2018). This detection was very surprising since no HF sferics were detected at Jupiter (probably due to ionospheric absorption, as pointed out by Zarka (1985)), and radio emissions of terrestrial lightning in the ultra-high frequency band (UHF, 300-3000 MHz) have rarely been studied due to the decline of intensity with increasing frequency. The MWR high frequency observations have been confirmed by parallel observations of whistlers with the Juno Waves instrument (Kolmasova et al., 2018; Imai et al., 2018). Modern receivers with low noise figures and wide bandwidth should allow good observations of impulsive radiation of lightning at microwave frequencies (Petersen and Beasley, 2014). Thus, the MWR lightning detections at Jupiter have opened up a new frequency window to study planetary lightning (e.g., at ice giants). At frequencies of a few hundred MHz the flux of Jovian synchrotron radiation from electrons trapped in the radiation belts is typically much higher than the flux from Jovian lightning (Brown et al., 2018), but Juno was flying below Jupiter's radiation belts, improving the signal to noise ratio. At Uranus there is no synchrotron radiation that could obscure potential microwave radio emissions from Uranus lightning.

Detecting optical emissions from lightning at Uranus and Neptune is probably very difficult, since the discharges might take place in the water or ammonium hydrosulphide clouds (see section 2.3) deeper in the atmosphere (40 x $10^3$ hPa or 40 bar) than at Jupiter or Saturn (Atreya and Wong, 2005). While many spacecraft easily detected the optical flashes from Jupiter's night side (Voyager 1 and 2, Galileo, Cassini, New Horizons, Juno), detecting optical flashes from Saturn's night side with the Cassini camera turned out to be more complicated. This was due to the ring shine and the greater depth of the discharges at the 8-10 x $10^3$ hPa level (Fischer et al., 2008) compared to typical depths of 5 x $10^3$ hPa at Jupiter (Dyudina et al., 2004). Finally, the first optical flash detection from Saturn's night side by Cassini was made around Saturn equinox in August 2009 when the ring shine was minimal (Dyudina et al., 2010). Interestingly, during the Great White Spot event on Saturn with its high SED rate of 10 $s^{-1}$ (Fischer et al., 2011), the Cassini camera also managed to image flashes on Saturn's day side with a blue filter by subtracting two temporally close images from each other (Dyudina et al., 2013). However, this technique might only work with high flash rates, and the UED rate measured by Voyager 2 was quite low. Nevertheless, optical images of atmospheric features at the ice giants are still very important since they can give clues about the location of possible lightning flashes or if storms are present at all. At Saturn, for example, it was found that storm clouds were brighter in the images when the SED rate was high (Dyudina et al., 2007), indicating enhanced vertical convection. Optical observations of the ice giants with ground-based telescopes or the Hubble Space Telescope are also important to study atmospheric dynamics and to specify times when it is worth searching for lightning radio emissions with large ground-based radio telescopes. We note that the LRD on-board the Galileo probe also had two photodiodes to measure optical flashes, but no optical signatures were found (Rinnert et al., 1998). Optical Jupiter lightning flashes were detected recently by the Juno orbiter's camera (JunoCam) and star tracker (Becker et al., 2019). If one does not intentionally fly into a thunderstorm (which would be very hard to realize technically at ice giants), detection of optical flashes or acoustic thunder with an in-situ probe seems improbable.

*2.2 Voyager 2 observations*

The PRA (Planetary Radio Astronomy) instrument on Voyager 2 detected 140 impulsive bursts in the frequency range of 0.9 to 40 MHz (upper frequency limit of the PRA) during the January 1986 Uranus flyby (Zarka and Pedersen, 1986). These bursts were termed UED (Uranian Electrostatic Discharges) in analogy to the similar radio emissions of SED (Saturn Electrostatic Discharge), detected by both Voyagers. The mean burst duration of the UED was 120 ms, and they were detected within distances of ~600,000 km of Uranus on $24_{th}$ -$25_{th}$ January 1986. Figure 1 shows both the UED rate as a function of distance in Uranian radii (1 $R_U$=25,600 km), and the distribution of all UED in the time-frequency plane. Due to the sweeping PRA receiver with a dwell time of 30 ms in each frequency channel, the UED are seen as short bursts over a limited frequency interval, despite the notion that they should be intrinsically broadband in reality. The low number of UED from ~20 to 30 MHz seen in the lower panel of Figure 1 is likely to be from an increased spacecraft noise level.

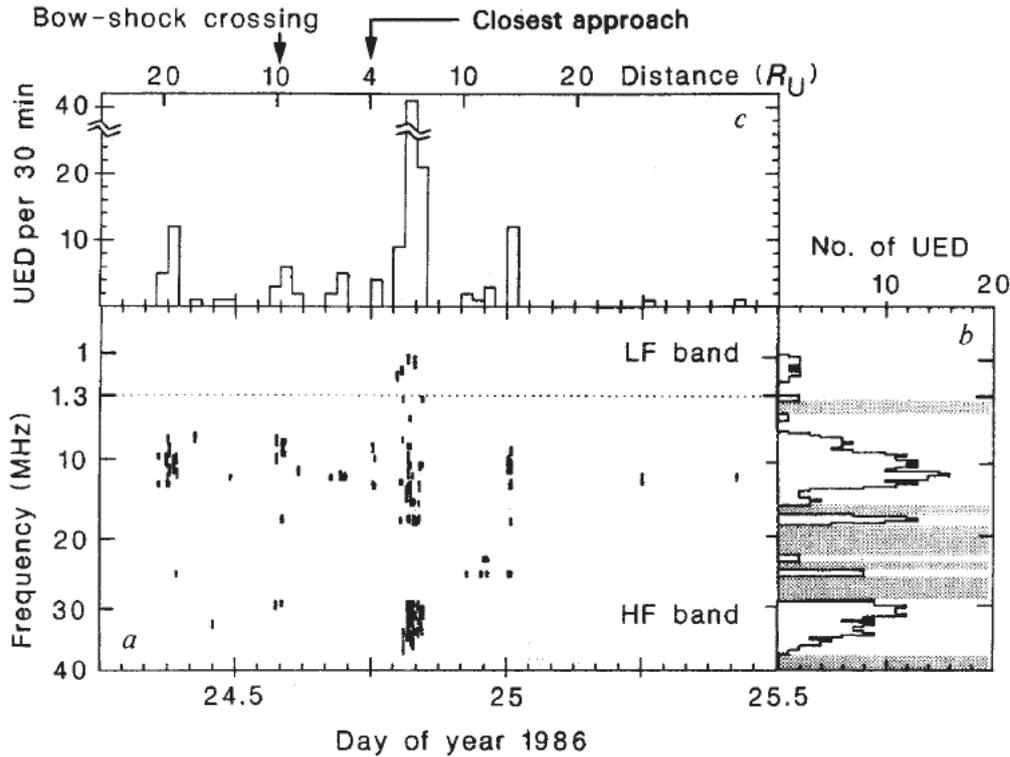

*Figure 1: Uranian Electrostatic Discharges detected by the Voyager 2 PRA instrument. Panel a (bottom) shows a dynamic spectrum, panel b (right hand side) the number of UED as a function of frequency and panel c (top) the number of UED as a function of time (Reproduced with permission from Zarka and Pedersen, 1986).*

Although the UED tend to group in episodes, no periodicity corresponding to the planetary rotation (~17.25 h) was detected, unlike SED. The intensity of UED is about an order of magnitude weaker than the intensity of SED. The average intensity normalized to the corresponding intensity that would be received at the Earth (at 1 AU) is $6\times10^{-24}$ W m$^{-2}$ Hz$^{-1}$ for the UED in the HF (high frequency) band (1.3-40 MHz) and $2\times10^{-22}$ W m$^{-2}$ Hz$^{-1}$ in the LF (low frequency) band (below 1.3 MHz). This corresponds to spectral source powers of 2 and 60 W Hz$^{-1}$, which is 2 to 3 orders of magnitude larger than the source power of terrestrial lightning, respectively. Neither whistlers nor optical signals of lightning or aurora were detected on the night side of Uranus by Voyager 2 (Smith et al., 1986).

During the Voyager 2 Neptune flyby on 25th August 1989, the plasma wave system (PWS) detected a series of 16 whistler-like events within ~20 minutes at radial distances from ~1.3 to 2 Neptune radii (1 $R_N$=24,762 km) and at magnetic latitudes from -7° to 33° (Gurnett et al., 1990). The frequencies ranged from 6 to 12 kHz, and the large dispersions around 26,000 sHz$^{1/2}$ fit the Eckersley law for lightning generated whistlers, for which the dispersion is frequency-independent (Rakov and Uman, 2003). Eckersley (1935) had shown that the arrival time $t$ of a terrestrial whistler is given by $t=t_0+D/sqrt(f)$ with $t_0$ as the time of the lightning flash, $f$ as the wave frequency, and $D$ as the dispersion constant. The dispersions are too large for a single direct path from the lightning source to the Voyager 2 spacecraft, and so the most likely propagation path involves lightning on the dayside of the planet with multiple bounces from one hemisphere to the other. Figure 2 shows a frequency time-spectrogram of Neptune whistler number 4, which lasts tens of seconds.

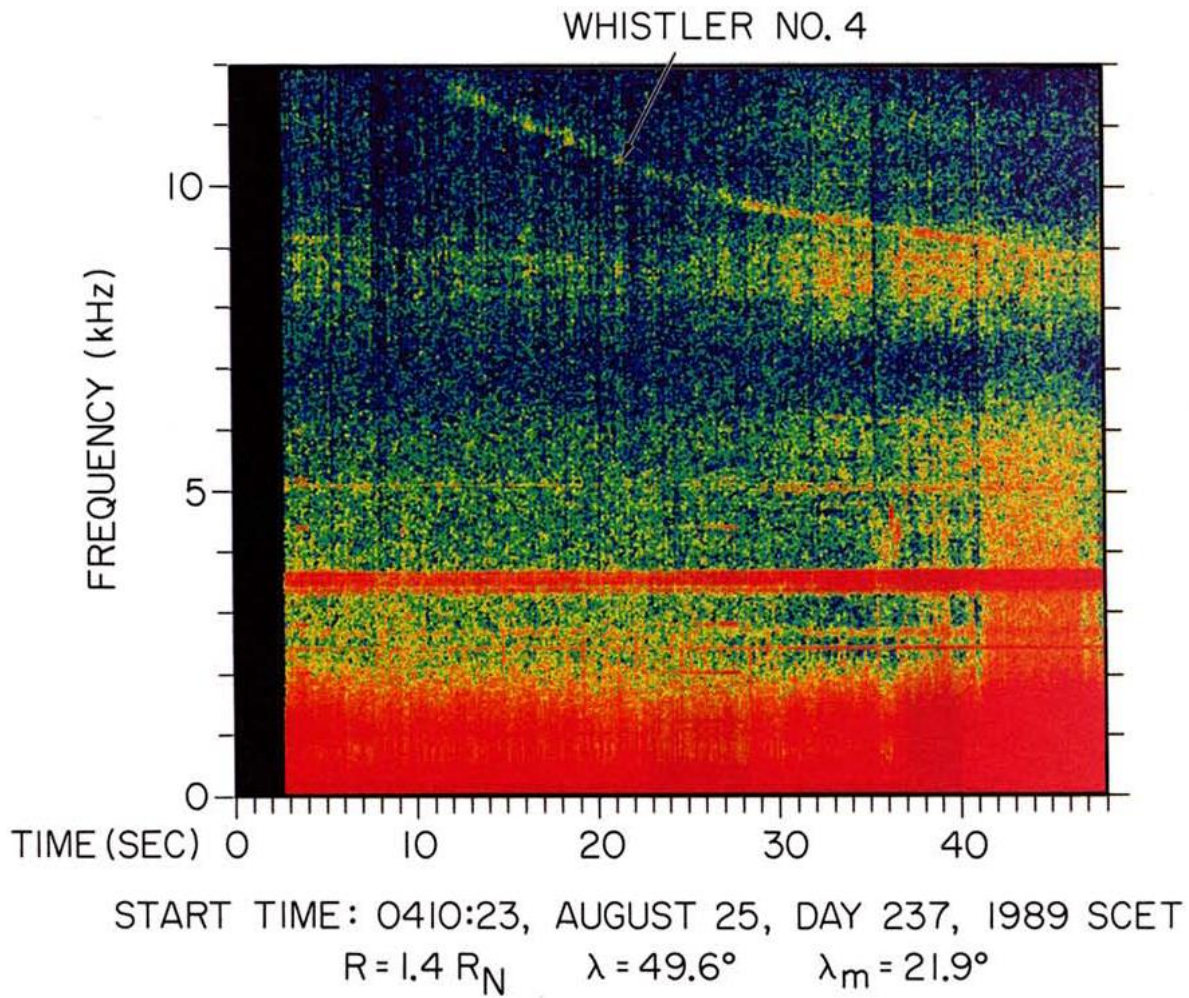

*Figure 2: Frequency-time spectrogram of a whistler recorded by the Voyager 2 plasma wave instrument at Neptune. The intensity is represented by the colour scale from blue (background intensity) to red (highest intensity). Reproduced with permission from Gurnett et al. (1990).*

Farrell (1996) interpreted the highly dispersed whistler-like signals as Z-mode radiation and not as whistler mode emission. Its source could be lightning, but a magnetospheric source is also possible. A magnetoplasma is a birefringent medium in which radio waves can propagate as ordinary or extraordinary waves. The Z-mode can be seen as the low frequency branch of the extraordinary wave, whereas the whistler is the low frequency branch of the ordinary wave (see, e.g., Gurnett and Bhattacharjee, 2017). The Neptune lightning hypothesis is somewhat supported by the fact that Kaiser et al. (1991) also detected four weak sferics at high frequencies (18-31 MHz) from a distance of 5-6 $R_N$ (Neptune radii) in the Voyager 2 PRA Neptune data. The average Neptune sferic intensity was $5 \times 10^{-18}$ W m$^{-2}$ Hz$^{-1}$ at 1 $R_N$ corresponding to an intensity of $\sim 1.35 \times 10^{-25}$ W m$^{-2}$ Hz$^{-1}$ at 1 AU, which is about 45 times weaker than the average UED intensity in the high band. The spectral source power of Neptune sferics would be $\sim 0.04$ W Hz$^{-1}$, which is comparable to the source power of a strong terrestrial lightning flash. No optical lightning detection was made by Voyager 2 at Neptune. We do not know if lightning on the ice giants is constant, like on Jupiter, or intermittent, like on Saturn.

Nevertheless, it is remarkable that Voyager 2 detected lightning at all four giant planets, albeit tentatively at Neptune. The properties of Uranus and Neptune lightning detected by Voyager 2 are summarised in Table 1.

*2.3 Possible origins of lightning - clouds and microphysics*

Uranus and Neptune have very similar atmospheric structures, inferred from remote sensing observations, radiative transfer and photochemical modelling. The most recent interpretation, broadly applying to both ice giants, (Mousis et al, 2018) has a stratosphere (0.1 - 30 hPa) of an extended, mainly hydrocarbon haze, generated by gravitational settling of aerosol particles from methane photolysis. In the troposphere there are expected to be ice cloud layers of methane ($CH_4$), with their base at 1300 hPa, a physically thin but optically thick hydrogen sulphide ($H_2S$) layer between 2000-4000 hPa, and beneath this ammonium hydrosulphide ($NH_4SH$), followed by water ($H_2O$) down to about $50 \times 10_3$ hPa. The water-ice cloud forms the top of a massive liquid water cloud that could extend down to at least $1,000 \times 10_3$ hPa (Mousis et al, 2018). In a study of Neptune cloud charging, a slightly different structure was assumed by Gibbard et al (1999). This included a region of ammonia ($NH_3$) ice cloud at the same level as the $H_2S$ ice cloud, with the deepest liquid cloud as a mixture of $H_2O$, $NH_3$ and $NH_4SH$.

Terrestrial thunderstorms are used as an analogy when considering whether these clouds could support lightning. Observations and experiments have shown that discharges are generated in mixed-phase water clouds, specifically, from collisional charge transfer between soft hail (graupel) and ice crystals, producing oppositely charged particles which are then separated by convection to generate a potential difference that eventually exceeds the breakdown voltage of air, causing a discharge (Saunders, 2008). Lightning at the giant planets has been attributed to a terrestrial-like process in mixed-phase water clouds, mainly because the flash depth from visible observations at Jupiter and Saturn is consistent with the anticipated depth and temperature range of the water cloud region (Aplin and Fischer, 2017). Lightning is possible in non-water clouds as long as there is adequate convection to create the clouds and sustain separation of the charged particles, and the cloud material is sufficiently polar to support charge transfer (physical properties of each of the proposed cloud layers are summarised in Table 2). Additional constraints related to the local atmospheric properties are that the breakdown voltage must be achievable by charge separation within the thundercloud. If the gas is too electrically conductive this limits particle charging through decreasing the relaxation time $\tau$ given by $1/\varepsilon_0 \lambda$ where $\lambda$ is the conductivity and $\varepsilon_0$ the permittivity of free space, and preventing an electric field from building up.

Gibbard et al (1999) simulated particle growth, charging, fall velocities and breakdown voltage for the cloud layers described above to determine which layer could support lightning, with collisional charging parametrised from laboratory experiments for terrestrial clouds (e.g. Saunders, 2008). $H_2S$ and $CH_4$ ice clouds were essentially ruled out as possible lightning generators due to their single phase and low polarisability. In the deep water cloud the limiting factor was the breakdown voltage, which is expected to be 250 MV/m at $50 \times 10_3$ hPa, whereas the electric fields achieved are only 10 MV/m. Electric fields are limited by electrostatic levitation of charged particles, which suppresses the generation of distinct areas of opposite charge within the cloud. Similar effects are expected in $NH_4SH$ clouds, but the electric field was a factor of 3 lower than the breakdown voltage. Based on this, Gibbard et al (1999) state that lightning is very unlikely in Neptune's water clouds, but could be possible in $NH_4SH$. These calculations were limited by a lack of data on the physical properties of $NH_4SH$, most likely because it is unstable at terrestrial surface conditions, hindering laboratory characterisation (Loeffler et al, 2015). Gibbard et al's (1999) results are consistent with the lack of optical detection of lightning from Uranus and Neptune, as lightning in the deep cloud layers would not be visible from orbit. This work also neglected the background conductivity of the gas in the cloud layers, for which no information was available (see section 3.3.1).

*2.4 Ground-based radio observations*

2.4.1 Lightning detection with ground-based radio telescopes

Searching for wide-band signals like lightning from other planets with ground-based radio telescopes is not a trivial task given the presence of Earth lightning and other natural and artificial radio interference. So far this has only been successful for Saturn (Zakharenko et al., 2012; Konovalenko et al., 2013), and we will describe in the following paragraph how this was done with the UTR-2 radio telescope.

The Ukrainian T-shaped Radio telescope model 2 (UTR-2) was constructed near Kharkov in the early 1970s, and it is still one of the largest ground-based radio telescopes in the decametric frequency range. The telescope is split into 12 sections that form three T-shaped arms (North, South, West) each 900 m long. In total it consists of 2040 fat linear dipoles (which have a broader frequency response than thin dipoles), with a frequency range of 8 to 32 MHz. UTR-2 has a large effective area of up to ~140,000 m² and a high directivity, with the main beam 0.5° wide (Konovalenko et al., 2016). UTR-2 can provide simultaneous observations with up to 5 spatially separated antenna beams, and the beam can be electronically steered within a wide range of both sky coordinates (azimuth, elevation). The multi-beam capability was essential for the detection of Saturn lightning, for which two beams were used, one directed at the source, Saturn, here called the ON beam, and one directed a few degrees off target (OFF). A Saturn lightning (SED) signal should only occur in the ON beam and not in the OFF beam, whereas most interference signals come in through the side lobes of the telescope and appear in both ON and OFF beams. The known characteristics of SED (duration, intensity, wide-band signal, almost flat spectrum in decametric frequency range) and the simultaneous SED observations with the Cassini Radio and Plasma Wave Science (RPWS) instrument (Gurnett et al., 2004) in the Cassini era (2004-2017) also helped to correctly identify the signals.

After the initial ground-based detection, SED were also observed with higher time resolution, and it was found that, just like the pulsars more usually observed with radio telescopes, the signals are dispersed by the interplanetary medium (and the ionospheres of Saturn and Earth) with a characteristic frequency-dependent propagation delay. This time is typically several hundreds of microseconds over a 10-20 MHz difference in frequency (Mylostna et al., 2013). This dispersion is typical in radio astronomy, and the time delay it causes can be defined in terms of the "dispersion measure" (DM), a constant which is expressed in units of parsecs per cubic centimetre (pc cm$_{-3}$), to represent the distance and the electron concentration in the interplanetary medium (e.g. Kraus, 1986). The DM is often found by empirically searching through a range of possible values to assess which gives the best overall signal-to-noise ratio. "De-dispersion" is often applied as a post-detection data analysis technique to compensate for the delay introduced by dispersion and maximise the signal-to-noise (e.g. Hankins and Rickett, 1975).

Zakharenko et al. (2012) suggested that the SED intensity peaks are in short bursts that become blurred at high time resolution. The dispersion delay across the range of frequencies observed would also affect the smoothing of short broadband bursts, especially if the bursts are infrequent. This was confirmed in high spectral resolution observations (Mylostna et al., 2014), Figure 3. An important benefit of ground-based SED observations is the discovery of several time scales in which Saturn's lightning was especially intense. In the case of the 2010-2011 storm (Fischer et al, 2011), these were characteristic durations of (a) tens of ms, (b) 30-300 µs, and (c) 2-5 µs (Mylostna et al., 2014).

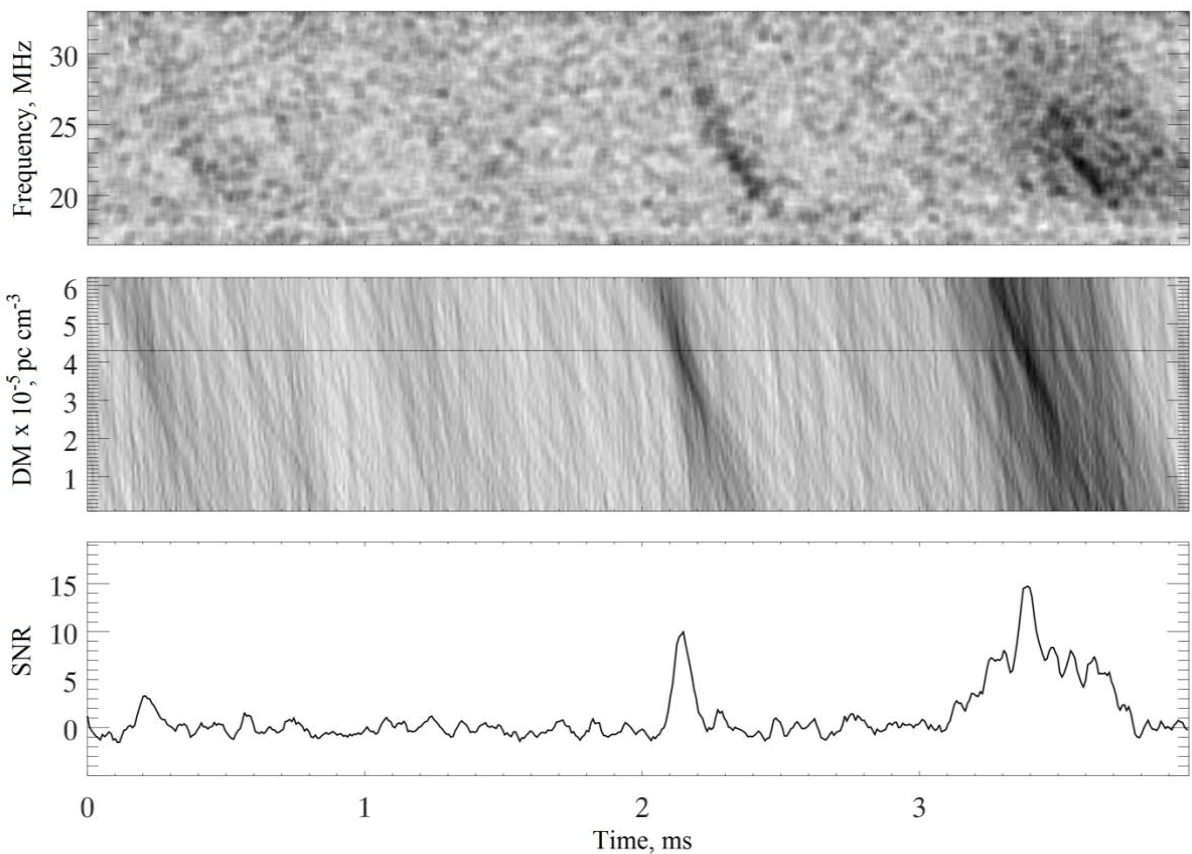

*Figure 3: Data processing of radio signals from SED starting at Dec 23 2010, 03h56m27.0s UT. Top panel shows the dynamic spectra of SED with a time resolution of 7 µs. The middle panel shows the same data, after application of a post-detection de-dispersion procedure, expressed in terms of dispersion measure (DM) in parsecs $cm^{-3}$ and with the maximum (43 $\times 10^{-6}$ pc $cm^{-3}$) indicated as a horizontal line. The optimal DM was found by manually searching from DM = (10 to 100) $\times 10^{-6}$ pc $cm^{-3}$ with a resolution of $10^{-6}$ pc $cm^{-3}$. The bottom panel shows the Signal-to-Noise ratio (SNR) at the optimal de-dispersion.*

Figure 3 shows that intense bursts only occupy a small fraction (10-20%) of the total flash duration. Therefore, their peak intensity when detected with a low temporal resolution will be significantly underestimated. In addition, the dispersion delay between the lower and upper frequency limits of 16.5 to 33.0 MHz is about 300 µs. Over the same period, the average duration of the most intense sub-millisecond components of the discharge ~70 μs. Thus, integration without eliminating dispersion delay also underestimates the lightning flux density. Figure 3 demonstrates this effect by showing the maximum flux densities obtained from the same data with and without elimination of the dispersion delay with a simple post-detector de-dispersion technique. The calculated flux density is enhanced by a factor of two if the de-dispersion is applied (Mylostna et al, 2014). The gain in sensitivity of a factor of two or three can be decisive for Uranus lightning detection, because without it the measurements are at the sensitivity threshold. In the next subsections, we will estimate this threshold in terms of the flux density of the UED (Zarka and Pedersen, 1986) and the use of radiometric gain. We will also discuss the possibilities of increasing sensitivity, using the radiative properties described above, optimising the observations for the presence of short bursts and dispersive delay of signals, and potentially with the help of two or more antennas far apart on Earth's surface.

2.4.2 Potential for ground-based observations of lightning from the ice giants

It will be shown below that the detection of Uranus lightning (UED) is within the technical capabilities of large ground-based radio telescopes (see also Zarka et al., 2004). The fluctuation $\sigma_{sky}$ of the galactic background is given by

$$4\sigma_{sky} = \frac{8k_B T}{A_{eff}\sqrt{\Delta f \Delta t}} \qquad (1)$$

where $k_B$ is Boltzmann's constant, $T$ the galactic background temperature, $A_{eff}$ the antenna effective area, $\Delta f$ the frequency bandwidth, and $\Delta t$ the integration time. We multiplied the sky background by a factor of 4 to account for the fact that a detectable signal should be at least a factor of 4 above the background fluctuations. The galactic background temperature is ~30,000 K at 20 MHz (see e.g., Kraus, 1986). The total effective area of the UTR-2 radio telescope is ~140,000 m², but here we take $A_{eff}$ = 90,000 m². This arises because for non-zenith sources, the effective area is scaled by cos($z$) where z is zenith angle. For example, for a source with declination = 0° and latitude of the UTR-2 = 49.63°, $A_{eff}$ ~ 90,000 m². In Figure 4 we have drawn $4\sigma_{sky}$ as a function of the receiver bandwidth (from 100 kHz to 10 MHz) and the integration time (20 ms or 0.1 s).

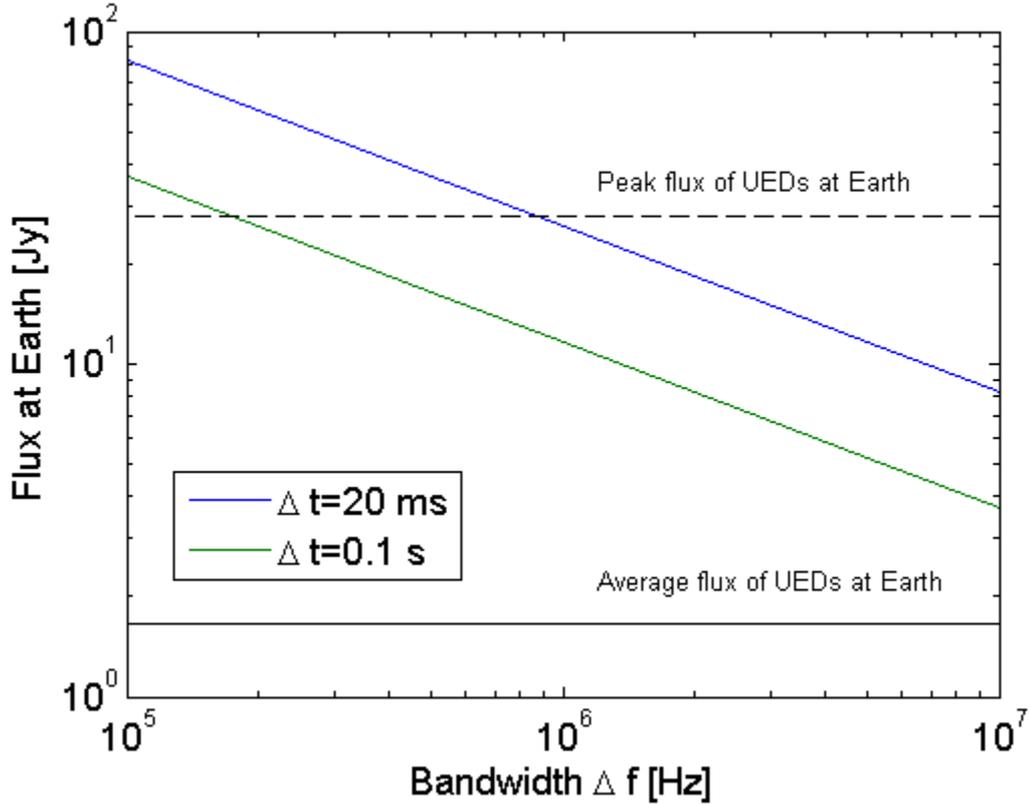

*Figure 4: Four times the galactic background fluctuation ($4\sigma_{sky}$) in Jansky (1 Jy = $10^{-26}$ Wm$^{-2}$Hz$^{-1}$) as a function of receiver bandwidth (100 kHz to 10 MHz) and integration time (blue line for 20 ms, green line for 0.1 s). The average and the peak flux of Uranus lightning (UED according to Zarka and Pedersen, 1986) at Earth are indicated by a solid and a dashed black line, respectively.*

Figure 4 shows that it is necessary to use at least a bandwidth of 1 MHz with an integration time of 20 ms to get a background fluctuation that is smaller than the peak flux of Uranus lightning (UED). The average UED flux at Earth was calculated using the flux of $6 \times 10^{-24}$ W m$^{-2}$ Hz$^{-1}$ in the HF band at 1 AU (Zarka and Pedersen, 1986), which translates to a flux of 1.7 Jy (1 Jy = $10^{-26}$ Wm$^{-2}$Hz$^{-1}$) at a distance of 19 AU (average Uranus-Earth distance). The peak flux of UED at Earth might be almost 30 Jy ($10^{-22}$ W m$^{-2}$ Hz$^{-1}$ at 1 AU around 15 MHz in Figure 4 of Zarka and Pedersen, 1986). Since the UED rate detected by Voyager 2 was low, one should base the choice of receiver bandwidth and integration time on the average UED flux, which is not even reached with a bandwidth of 10 MHz. An integration time of 0.1 s is of the same order as the expected signal duration, which is a reasonable choice to achieve a first detection. Longer integration times would dilute the signal, and shorter integration times would need strong UED around the peak flux which should be rather rare events. It is important to note that for short signals like lightning one cannot simply enhance the detectability by using very long integration times. With the UTR-2 frequency range of 8-32 MHz (Konovalenko et al., 2013) one also cannot have a much larger bandwidth either. The integration in bandwidth can be done in the post-processing stage, so it is possible that the receiver bandwidth during the actual observation is smaller. The same holds for the integration time.

We conclude that UED detection should be possible with the UTR-2 radio telescope, but we are close to its sensitivity threshold. In contrast to the UED, the SED intensity at Earth are a few hundred Jy on average with a peak intensity as high as 45000 Jy. This has enabled study of the fine structure of SED down to the microsecond range (Mylostna et al., 2014). Finally, we note that the expected average flux of Neptune lightning at Earth would only be around 15 mJy (Kaiser et al.,1991), which would need a radio telescope more than 100 times larger than UTR-2 for a detection in the decametric frequency range.

2.4.3 First ground-based attempts at Uranus lightning detection

In summer 2014 ground-based infrared images made with the W.M. Keck observatory showed several storms in the atmosphere of Uranus. In spite of the expected decline in convective activity following the 2007 equinox, eight storms were detected on the planet's northern hemisphere on August 5-6 2014 (de Pater et al., 2015). One of them was the brightest storm ever seen on Uranus, located around a planetocentric latitude of ~15°N and reaching altitudes of ~330 hPa, well above the uppermost methane-ice cloud layer. The brightness of this feature had already decreased substantially by August 17, and it might have been formed by strong updrafts. Another, deeper, cloud feature (at about 2000 hPa) was seen later (October 2014) by amateur astronomers and by the Hubble Space Telescope at a latitude of 32°N, but overall the storm activity was significantly decreased by October 2014.

Based on initial information from infrared and optical observations, two campaigns were conducted at UTR-2 in 2014: August 18-25 and October 6-12. As previous work had indicated that the source of lightning may not be tied to the exact position of the storm, observations were made during the entire period when the planet was above the horizon and the effective antenna area did not drop very much. Observations with time interval +/- 3 hours from culmination give a zenith angle of Uranus in culmination ~45° (declination of the planet in August-October 2014 was about 5°), and near 70° at the start and end of a measurement sequence. The observation technique was as follows: three receivers in correlation mode (Zakharenko et al, 2016) of antennas North-South and West-East (which provides the maximum set of analyzed parameters: module and phase of antenna signal cross-spectra and their individual power spectra) were connected to beams 1, 3 and 5 of the radio telescope. Beam 3 was directed at the source (ON), and beams 1 and 5 (both OFF) were turned away from the source by 1° along the meridian: beam 1, to the south and beam 5 to the north. The height of the source above the horizon varied from 20 to 45 degrees, while the effective area of the radio telescope was 50,000 - 100,000 m2. With a bandwidth of about 10 MHz and an integration time of 20 ms, the sensitivity of the UTR-2 was sufficient to detect the maximum lightning flux (see Figure 3). However, over 15 days of observation, there were no events that were clearly visible in the ON beam and absent in the OFF beams.

Subsequently, one week of similar Uranus observations have been carried out each September-October since 2015, when the culmination of Uranus in the middle of the night provided the minimum radio frequency interference and therefore the best conditions for scanning observations. No lightning signals from Uranus have yet been recorded.

3. Ionisation and particle charging

Ions are present in all planetary atmospheres, and electrons are present where chemistry permits, making the air a weak conductor of electricity. These ions and electrons interact with atmospheric clouds, dusts or hazes (all referred to here as "aerosol", a particle suspended in a gas) to attach to, and charge them, meaning that some fraction of atmospheric aerosols are charged, with their charge obeying a Maxwell-like distribution (e.g. Gunn, 1954). In this section we outline the physics of ionisation and ion formation, how ions and electrons interact with aerosols and the consequences for weather and climate of the ice giants.

*3.1 Sources of ionisation*

As was explained in Section 1, galactic cosmic rays (GCR) are the most penetrating source of ionising radiation in planetary atmospheres. Other ionising radiation in the Solar System includes natural radioactivity, mainly emitted from the surfaces of rocky bodies and so not considered further here. Photoionisation from UV is relevant for ice giant stratospheres but not tropospheres, due to absorption by stratospheric haze (Moses et al, 1992). Photoemission of electrons from aerosols has been considered for other planetary atmospheres such as Titan, (Whitten et al, 2008) but is also assumed not to occur in the ice giant troposphere due to the lack of UV radiation. Moons often receive a flux of energetic electrons from the magnetospheres of the planets they orbit, which provides an additional source of ionisation for the tenuous atmosphere of Neptune's moon Triton (e.g. Delitsky et al, 1990) and will be discussed further in section 3.5.

GCR are typically energetic protons and alpha particles created by energetic astrophysical events, such as the shock fronts of expanding supernovae remnants (Blandford and Eichler, 1987; Hillas, 2005). Incident GCR propagate through planetary atmospheres until the point at which they experience an inelastic collision with an atmospheric nucleus. This inelastic collision leads to a secondary particle cascade, whose flux continues to build until the so-called Pfotzer-Regener maximum is reached, after which the flux of secondary particles (and resulting atmospheric ionization) begins to decay with increasing atmospheric pressure. The most energetic particles are muons created from pion decay which, as on other planets such as Earth, can ionise the deep troposphere (e.g. Aplin, 2013). GCR ionisation is therefore considered to be the only source of tropospheric ionisation at Neptune and Uranus, with UV also contributing in the stratosphere.

*3.2 Modulation of ionisation*

3.2.1 Heliospheric magnetic fields

Planetary ionisation is modulated inversely by the 11-year solar cycle, due to the Sun's magnetic field deflecting GCR away from the Solar System more strongly at solar maximum, so the GCR flux is generally anticorrelated with the solar UV flux. Lower-energy GCRs are affected proportionally more by the solar cycle, which has consequences for atmospheric ionisation, as the lower-energy GCRs are more likely to lose their energy at relatively high altitudes. For example, Nordheim et al. (2015) showed that for Venus, another deep planetary atmosphere, the difference in ionisation rate between solar maximum and minimum is negligible below the tropopause due to the dominant contribution from energetic particles with little solar modulation. At the ice giants, GCR shielding due to planetary magnetic fields will preferentially lead to differences in ionisation rate at higher altitudes in addition to the effect of magnetic latitude.

3.2.2 Planetary magnetic fields

Planetary magnetic fields deflect GCR, resulting in a latitudinal variation where lower-energy primary GCR can enter atmospheres at the magnetic poles, but only higher-energy particles can enter near the magnetic equator. Most planetary magnetic field axes are closely aligned with their geographic spin axes, but Neptune and Uranus are different. If the magnetic fields are modelled as a simple dipole (i.e. a bar magnet inside the planet), then the spin axis-dipole tilt of Uranus is 59° and of Neptune 47° from their respective axis of rotation, with the effective dipole centres (i.e. the bar magnet itself) offset from the centre of the planet, by a larger amount for Neptune than for Uranus (Nellis, 2015). This means that the variation of ionisation rate with geographic latitude is not similar to geomagnetic latitude, as for Earth, and will be asymmetric across the planet's hemispheres. Both magnetic fields are similar in magnitude to Earth's, with Uranus slightly greater in magnitude than Neptune (Nellis, 2015), implying that Neptune will have a greater ionisation rate since a larger fraction of the GCR spectrum can access its atmosphere.

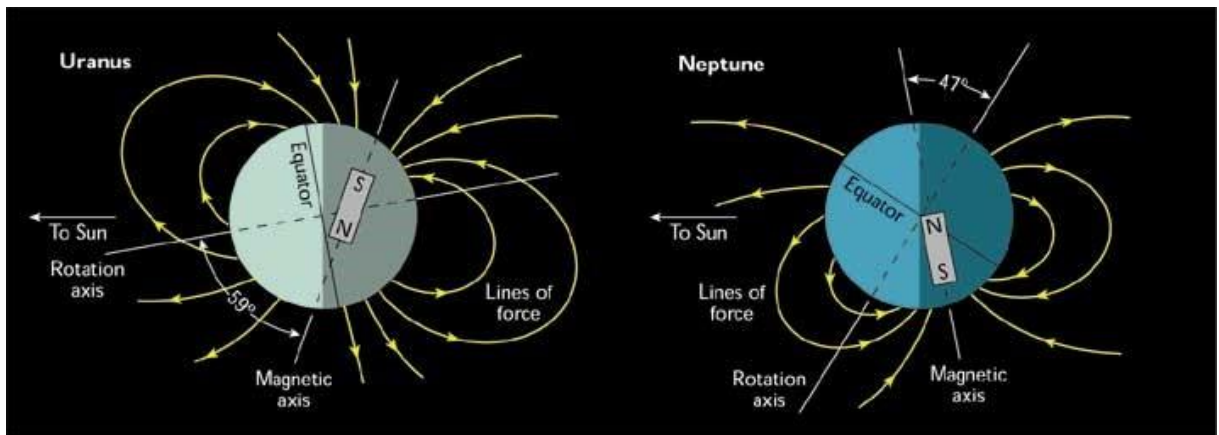

Figure 5: The magnetic fields of Uranus and Neptune as measured by Voyager 2 (image provided by NASA)

3.3 Atmospheric ions

Ionisation in planetary atmospheres creates a primary ion and an electron, which then react chemically to form clusters whose composition depends on other species present, and their hydrogen and electron affinities (Aplin and Fischer, 2019). Capone et al (1977) predicted the terminal, most abundant positive ice giant cluster-ions to be $CH_{5+}(CH_4)_n$ (with $n$ = 1 or 2 most commonly), with negative particles expected to stay as free electrons due to the lack of electrophiles. However, this pre-Voyager study was limited to the stratosphere. Moses et al (1992) took a similar approach with simplified chemistry using atmospheric data from Voyager 2, with the main difference from the Capone et al (1977) model being that some of the $CH_4$ ligands were replaced by other condensable hydrocarbons in the stratosphere.

The presence of atmospheric ions makes the air slightly electrically conductive, with the (positive or negative) conductivity $\lambda$ related to the mean ion and/or electron concentration $n$ and mobility $\mu$ by $\lambda = nq\mu$, where $q$ is the charge on the electron. Mobility defines the speed of a particle in a unit electric field and is related its mass, the ambient gas and its local properties.

Mobility is related to mean free path, so it increases with temperature but decreases as atmospheric pressure increases. A linear assumption is commonly used to calculate pressure and temperature effects on mobility, but its linear variation with temperature has been challenged. Mobility is also greater in atmospheres with a less massive background gas (Harrison and Tammet, 2008).

In atmospheres containing free electrons, the negative conductivity exceeds the positive conductivity by several orders of magnitude due to the electron mass, which is orders of magnitude smaller than that of a cluster-ion. For example, Titan's atmosphere was expected to contain free electrons, with considerable uncertainty on the prevalence of electrophilic species. Conductivity measured by the Huygens probe was lower than predicted, indicating that more electrophilic species and fewer free electrons were present (e.g. Aplin, 2013). (As an aside, this is an example of how in situ electrical measurements can be used to constrain atmospheric composition). The ratio between positive and negative conductivity determines the rate at which positive and negative charged particles attach to aerosol, which will be discussed in section 3.4.

3.3.1 Estimating air conductivity

There is no ion-aerosol model for the ice giant atmospheres, but it is possible to estimate the atmospheric electrical conductivity around the tropopause, which is usually close to the maximum ionisation rate (see section 3.1). The tropopause is at a temperature of 55K, at 200 hPa on Neptune and 160 hPa on Uranus. Previous models (Capone et al, 1977; Moses et al, 1992) assumed no electrophilic species, but more recently, electrophilic trace species in the stratosphere such as $CO_2$ and $H_2O$ have been identified above the 100 hPa layer (Mousis et al, 2018), as well as the non-electrophilic tropospheric trace species $PH_3$ (Teanby et al, 2019) and $H_2S$ (Irwin et al, 2019). Negative ions created by electron attachment to electrophiles would need to be included in any future model, particularly as the trace species appear quite different between Uranus and Neptune. Here, electrophiles are neglected due to lack of data, with negative conductivity assumed to be from free electrons only.

Electron mobility in hydrogen and helium at 77 K and 200 hPa $\sim 2$ $m^2V^{-1}s^{-1}$ (Pack and Phelps, 1961; Ramanan and Freeman, 1990, 1991). A typical ice giant atmospheric electron mobility can be obtained from a weighted average of the slightly different fractions of hydrogen and helium at each planet. A scaling factor was given by Harrison and Tammet (2008), indicating that for ions of equal mass, mobility in hydrogen at 100K would be a factor of 4.5 times greater than in nitrogen at the same temperature. Assuming positive cluster-ions at the ice giants are roughly the same mass as terrestrial cluster-ions, i.e. a few tens of atomic mass units, then this scaling factor of 4.5 can be applied, in combination with a linear scaling for atmospheric pressure, to estimate the mean mobility with respect to cluster-ion mobilities at the terrestrial surface ($\sim 10^{-4}$ $m^2V^{-1}s^{-1}$). Finally, modelled ion and electron concentrations from Capone et al (1977) near the tropopause, of $10^4$ $cm^{-3}$ are assumed. The results, shown in Table 3, indicate that as anticipated, the negative conductivity dominates due to the presence of free electrons. The initial estimates in Table 3 can be used as a basis on which to specify instrumentation, which clearly needs to have a wide bipolar range to deal with the significant conductivity asymmetry.

The likelihood of lightning is related to the atmospheric conductivity. Michael et al (2009) argued against lightning on Venus on the basis that the atmosphere was conductive enough, and the breakdown voltage large enough, that the charging rate would never be sufficient for the breakdown voltage to be reached. Applying a similar argument, the ice giant atmospheres have lower breakdown voltages, but more conductive atmospheres due to their free electrons. With charge separation inhibited, as described in Section 3, it is difficult to understand which regions of the atmosphere are most likely to support lightning without more detailed modelling work.

*3.4 Ion-aerosol interactions at the ice giants*

Atmospheric ions and electrons attach to aerosol and transfer their charge. This reduces the number of ions and electrons, which in turn reduces atmospheric conductivity, whilst shifting the space charge to larger particles, which acquire a charge distribution. Charge on aerosol can affect coagulation and lifetime, as described by ion-aerosol theory (Gunn, 1954), and in combination with feedbacks and other processes can ultimately lead to meteorological effects on optical depth, visibility, clouds and precipitation. Ions themselves may also grow to become charged aerosol particles; this will be discussed in section 3.5.1.

The high mobility of free electrons means that their attachment to any clouds or hazes can be significant. The importance of charge in planetary atmospheres was first identified by Toon et al (1980) who recognised that photoelectric charging from UV radiation would release electrons in the upper atmosphere of Titan, and that this charge would be relevant to coagulation. Subsequent modelling and measurements have revealed the significance of charge for Titan's haze (e.g. Aplin, 2013). This modelling approach has recently been developed to consider the Uranus stratospheric haze (Toledo et al, 2019), where it is shown that equilibrium timescales for particles of 0.1-0.3 $\mu$m are enhanced by up to an order of magnitude by the presence of 10 elementary charges per $\mu$m of radius per particle.

Despite the clear significance of charge in the ice giant atmospheres, there have not been any studies of ion-aerosol physics in these environments, with charge parameters in models necessarily based on simple estimates. Ion-aerosol interactions have, however, been modelled for the atmosphere of Jupiter (Whitten et al, 2008). The similarities between the gas and ice giants, particularly the hydrogen and helium atmosphere, and the presence of free electrons, means that some of these findings might apply at the ice giants.

The Whitten et al (2008) Jupiter study considered three monodisperse cloud layers down to a pressure of 5.5 x $10^3$ hPa, above the liquid water clouds, with GCR as the ionisation source. Ion/electron losses by recombination and attachment to aerosol were modelled using a set of coefficients related to the number and size of particles and their charge. The key finding was that there were few free electrons due to "collection" by cloud particles. Positive ion concentrations were enhanced compared to cloud-free air, since there were fewer electrons available for recombination. The clouds had a bipolar charge distribution, with +6q (where q is the number of elementary charges) as the most likely charge state for any one particle, but a net negative charge overall with some particles carrying up to -30q.

Similar effects can be anticipated in the ice giant clouds, as long as the aerosol number concentrations are not significantly lower than the Jovian values of about $10^{10}$ m$^{-3}$. Converting the cloud estimates presented in grams per litre in Mousis et al (2018) to particles per cubic metre requires knowledge of the particle size, which is only available for the methane ice cloud (0.1-0.2 $\mu$m) and the deep water cloud (1-1.5 $\mu$m). Assuming the density of $CH_4$ ice to be 430 kg $m^{-3}$ (Satorre et al, 2008), the number concentration of $CH_4$ ice cloud is estimated to be $10^{16}$ $m^{-3}$ and the deep water cloud $10^{15}$ $m^{-3}$. These estimates have high uncertainty due to the assumption of sphericity, and sensitivity to cloud particle size, which is based on degenerate retrievals from radiative transfer models, but they are clearly greater than the Jupiter cloud particle concentrations. The clouds on Neptune and Uranus are therefore expected to be net negatively charged and with few free electrons in the cloud layers. The lower temperatures at the ice giants are not expected to affect this result, since the electron mobility will still significantly exceed the positive ion mobility.

As indicated above, the electrical properties of the stratospheric haze are also likely to be significant. There will be charging from photoemission as well as GCR, and negative ions are also likely due to the electrophilic trace species. More detailed investigations are necessary to better constrain charge effects in the haze.

*3.5 Solar cycle variations and ion-induced nucleation*

3.5.1 Neptune and Uranus

In long-term ground-based observations of Neptune at two wavelengths from a 21-inch telescope at Lowell Observatory, Arizona, Lockwood and Thompson (2002) demonstrated that the astronomical magnitude, representing the disk-averaged brightness, showed an 11-year periodicity consistent with the solar cycle, once seasonal fluctuations in the brightness had been removed. Aplin and Harrison (2017) used detrending based on robust fitting techniques to show a similar 11-year periodicity in brightness observations of Uranus made with the same telescope (Lockwood and Jerzykiewicz, 2006).

This solar cycle variation can have two possible causes, related to UV or GCR. Baines and Smith (1990) suggested a "tanning" mechanism for Neptune, where UV radiation modified the colour of particles, whereas Moses et al (1992) proposed that solar variation could be accounted for by ion-induced nucleation onto ions formed by GCR. (Ion-induced nucleation is a process where gases condense onto ions to create small particles that can ultimately act as cloud condensation nuclei). In a statistical analysis, Aplin and Harrison (2016) found that both UV and GCR mechanisms explained the observations. Over the duration of the observations (1972-2014), UV was the most likely mechanism for the solar cycle at 472 nm, accounting for 20% of the variance, but for 551 nm a combination of UV and GCR was required to provide the best explanation. Aplin and Harrison (2016) also used a known spectral "fingerprint", uniquely found in GCR data, which has previously been used to distinguish solar irradiance from GCR effects in Earth's atmosphere (Harrison, 2008). The "fingerprint" periodicity was particularly strong in the 1980s, and could be used with GCR measured both on Earth and by Voyager 2 to demonstrate that the GCR fingerprint was present in Neptune's brightness fluctuations, when Voyager 2 was close to Neptune. During the 1980s both wavelengths were statistically significantly explained by GCR variations, but the 472 nm wavelength was most responsive to GCR, indicating that the dominant mechanism can change over time.

In a follow-up study, Aplin and Harrison (2017) identified an 11-year periodicity in the Uranus observations for the first time through spectral analysis. Statistical analysis revealed a stronger solar signal with UV and/or GCR explaining up to 24% of the variance in brightness fluctuations, compared to 20% for Neptune. GCR effects also seemed more prominent on Uranus than on Neptune, both in terms of the statistics of fitting to the different physical models, as summarised in Table 4, and the change in brightness per unit change in GCR flux. At Uranus for 551 nm the normalised response to GCR was 0.07±0.02 units of astronomical magnitude per fractional change in GCR flux whereas at Neptune it was 0.04±0.04.

Table 4 also summarises the likely atmospheric origin of the disk-averaged brightness fluctuations, through estimating the cloud type at which the optical depth is 1. These visible wavelengths are dominated by the troposphere, on Neptune in regions corresponding to $CH_4$ ice cloud, and on Uranus to $H_2S$ ice cloud. The $CH_4$ cloud observations are consistent with Moses et al (1992) who predicted that the effects of ion-induced nucleation would be detectable due to GCR variations across the solar cycle. Methane's triple point is only 10K above the temperature at the condensation level, allowing it to nucleate as a supercooled liquid and then freeze (Moses et al, 1992). This seems a more likely mechanism than ions acting as centres for the nucleation of ice particles, for which there is little evidence (Seeley et al, 2001).

In Uranus the signals at 472 and 551 nm appear to originate from deeper in the troposphere, coinciding with a layer of $H_2S$ ice cloud. Although the troposphere is relatively inaccessible to remote sensing, classical physical theory can be used to estimate the supersaturation of $H_2S$ needed for ion-induced nucleation (Aplin, 2006, Moses et al, 1992). This is essentially the excess "relative humidity" of $H_2S$ required with respect to the air, before the gas begins to condense out onto ions or other nuclei to make small particles, described by:

$$\ln S = \frac{M}{k_B T \rho}\left[\frac{2\gamma_T}{r} - \frac{q^2}{32\pi^2 \varepsilon_0 r^4}\left(1 - \frac{1}{\varepsilon_r}\right)\right] \qquad (2)$$

where $S$ is the supersaturation, $M$ the molecular mass, $k_B$ the Boltzmann constant, $T$ the temperature, $\varepsilon_0$ is the permittivity of free space, $\varepsilon_r$ is the relative permittivity, $\gamma_T$ the surface tension, $\rho$ the density, and $q$ the charge on the electron. This equation describes the maximum supersaturation that is needed for a droplet of radius $r$ to exceed the energy barrier for nucleation. Here estimates for $H_2S$ at 188K are used, with $\varepsilon_r$ =10.487 (Harvey and Mountain, 2017), $\gamma_T$ = 0.0388 N/m (Riahi and Rowley, 2014) and $\rho$ = 993 kg m$^{-3}$ (Greenwood and Earnshaw, 1997). The results for the top (90K) and bottom (120K) of the cloud layer for a range of electronic charges are shown in figure 6. The maximum at each charge level indicates the supersaturation required for nucleation at the "critical radius". The maximum charge theoretically possible on a droplet of $H_2S$ before instability sets in, the Rayleigh limit, (e.g. Schweizer and Hanson, 1971) is also shown for comparison.

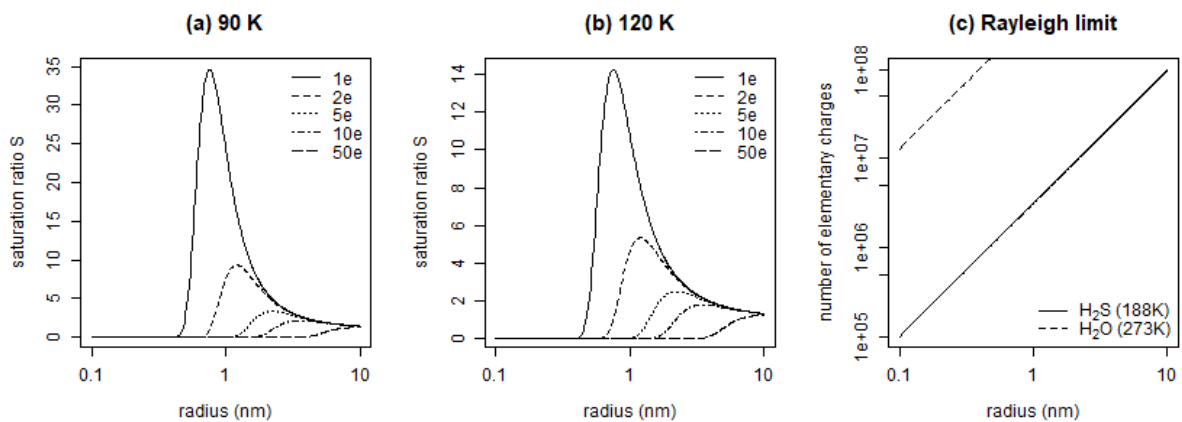

*Figure 6: Saturation ratio needed for condensation of $H_2S$ onto ions with between 1 and 10 elementary charges at temperatures corresponding to the Uranus cloud (a) top and (b) bottom with ( c) showing the maximum number of charges that can be sustained on $H_2S$ droplets (solid line), with $H_2O$ at 273 K for comparison (dashed line).*

The actual supersaturation is not known, but is suggested by Irwin et al (2018) to be 0.13±0.12, which would require a charge of 50q on a 7nm particle at the cloud base for nucleation. Charges of 30-50q on small aerosol particles have been predicted in the electron-rich atmospheres of Jupiter (Whitten et al, 2008) and Titan (Molina-Cuberos et al, 2018), and are well under the limit shown in figure 6(c). However, there are many uncertainties due to a lack of laboratory data on $H_2S$ and the difficulties of remotely sensing the troposphere. It is particularly unclear how ion-induced nucleation could contribute to the formation of ice cloud, as the temperature in the $H_2S$ clouds is much cooler than its triple point of 187K (Goodwin, 1983). Freezing of pre-formed liquid $H_2S$ droplets lofted by convection from warmer regions is one possibility, consistent with the stronger role for GCR at the bottom of the cloud identified by the statistical study of Aplin and Harrison (2017). Interestingly, Irwin et al (2018) suggested that the cloud particle albedo is consistent with the presence of photochemically formed products drizzled down from the stratosphere. This could potentially explain the role for UV in the upper parts of the tropospheric cloud implied in the statistical modelling summarised in table 3.

3.5.2 Triton

Voyager 2 unexpectedly observed weather, in the form of fogs, clouds and hazes, in the thin atmosphere of Triton, Neptune's largest moon. Triton's atmosphere is mainly nitrogen with a surface pressure of only $10^{-3}$ hPa and temperatures of approximately 40K. Delitsky et al (1990) suggested that nitrogen ion clusters would be abundant, with Neptune's magnetosphere as the dominant source of radiation, and additional ionisation from GCR and UV. Based on the likelihood of high supersaturation with respect to nitrogen, and the stability of large ion clusters close to the critical threshold for nucleation, it was predicted that the clouds and hazes at around 9 km altitude could be created by ion-induced nucleation. However, it remains unclear how ions can assist in nucleation at very low temperatures. Triton's atmospheric temperatures are well below the freezing temperature of nitrogen, but the classical cloud physics theory outlined in section 3.5.1 above is only for gases condensing on liquid drops. Ice also needs to nucleate onto something, and there is neither theoretical nor experimental support for ions acting as ice nuclei (e.g. Seeley et al, 2001). It seems more likely that there are alternative sources of aerosol in Triton's atmosphere, such as photochemistry (e.g. Zhang and Strobel, 2018). There has been very little theoretical or experimental work on ion-induced nucleation of liquids or ices beyond terrestrial conditions; more would increase our understanding of ion-induced nucleation in planetary atmospheres.

## 4. Discussion

*4.1 Past observations and interpretation*

The suggestive observations of electrical discharges by Voyager 2 at both Uranus and Neptune in the 1980s indicated that of the two planets, Uranus was more electrically active with both stronger and more frequent radio emissions. Although Uranus was mostly featureless in the Voyager visible light observations, it has subsequently been revealed to have a dynamic and rapidly evolving atmosphere with active convective storms. Uranian lightning is expected to be strong enough to be detectable from Earth with a large radio telescope, but an observation campaign during and after the 2014 intense storm period did not observe any discharges. In section 4.3 below the prospects for future ground-based lightning detection are discussed. Microphysical modelling suggests that lightning on both planets is generated in the deep troposphere, although it seems to be more likely in the ammonium hydrosulphide cloud rather than the water cloud. This would explain the lack of visible detection of lightning in comparison to the gas giants, where the water clouds are less deep.

Analysis of a long time series of telescope observations of the disk-averaged brightness of the ice giants demonstrated that both UV and GCR were modulating tropospheric brightness fluctuations, probably in the $CH_4$ (Neptune) and $H_2S$ (Uranus) ice clouds. This is the first identification of a significant role for ion-induced nucleation in a planetary atmosphere, with a common modulation by the host star. Cloud and aerosol measurements, combined with electrical properties, would provide more information on this solar modulation. Basic information on the chemical properties of species in the ice giant atmospheres is sparse, particularly for $NH_4SH$ and $H_2S$; more laboratory measurements of their properties at temperatures <100 K are needed to improve understanding of ice giant cloud microphysics.

*4.2 Recommended technologies for future missions*

Lightning is detectable from orbit, whereas the so-called "fair weather" atmospheric electrical properties require in situ instrumentation such as a probe. A minimum payload for any progress in ice giant atmospheric electricity would be a radio antenna on an orbiter, similar to that carried by Voyager 2.

4.2.1 Atmospheric electricity instrumentation

An ice giant descent probe should contain an atmospheric structure instrument, similar to the one suggested for the Hera Saturn entry probe (Mousis et al., 2016) or the Huygens HASI instrument (Fulchignoni et al, 2005). It should consist of an atmospheric electricity package in addition to an accelerometer, a temperature and a pressure sensor. Lightning should be detectable by a short electric or magnetic antenna (monopole, dipole, loop or spherical double probe) with a corresponding receiver in the VLF range where the signals are expected to be most intense.

The conductivity of the atmosphere can only be measured in situ. Mutual impedance probes send a current pulse through the surrounding medium, and the impedance can be determined from the current/voltage characteristic measured by two passive electrodes. A relaxation probe can also measure conductivity, and the spectrum of ion mobilities (Aplin, 2005), from the rate of decay of the potential on an electrode. Given the wide bipolar conductivity range throughout the ice giant atmospheres, and the sensitivity of the negative conductivity to the poorly-known number of electrophiles, a wide-range instrument package is recommended similar to the Pressure Wave Altimetry (PWA) package on the Huygens probe. On the PWA a mutual impedance probe was sensitive to conductivities $10^{-11}$-$10^{-7}$ S/m and two relaxation probes covered $10^{-15}$-$10^{-11}$ S/m (Molina-Cuberos et al, 2001). The three instruments provide redundancy in the event of instrument or data transmission failure, as happened for the Titan descent. Conductivity instruments can be used to estimate the number concentration of ions and electrons and their mobility (Aplin, 2005), thus providing clues to atmospheric composition. In combination with other instruments like nephelometers, conductivity can permit calculation of cloud and aerosol particle charging. The probes from relaxation instruments can also be used to measure DC electric field to deduce cloud and aerosol properties and even the existence of a global electrical circuit (Aplin, 2006).

For measurement of sferics and whistlers from lightning discharges an orbiter should be equipped with a radio and plasma wave instrument, capable of measuring signals at least from the VLF to the HF range (3 kHz to 30 MHz). This would enable measurement of periodic auroral radio emissions and various plasma waves as well as whistlers and sferics. Due to the large tilt of the magnetic field axis with respect to the rotational axis (see section 3.2.2) the ice giant magnetospheres should be highly dynamic, and their investigation should be a prime scientific objective of any mission. The instrument design can strongly benefit from the heritage of the Cassini RPWS (Radio and Plasma Wave Science) instrument (Gurnett et al., 2004), especially its gonio-polarimetric capability (Cecconi and Zarka, 2005). An additional feature could be waveform receivers with a sampling time of the order of microseconds, which could resolve the sub-strokes of a lightning flash. A millisecond mode, as realized with Cassini RPWS, would not suffice for this task. Due to memory and telemetry restrictions such a waveform receiver can only take short snapshots, and it should have a trigger system to eliminate "empty" snapshots with no signal.

A microwave radiometer (MWR) on an ice giant mission can be a versatile instrument. Besides its main task of investigating the dynamics and composition of the atmosphere down to pressure levels of several hundred bars (Janssen et al., 2017), the Juno MWR can also be used for lightning detection (Brown et al., 2018) if there is suitably low noise and a large bandwidth. Since each mission to an ice giant will most likely have a camera system, it can be used to search for optical flashes, although they might not be easy to find as pointed out in section 2.1.

*4.3 Future ground-based searches for Uranus lightning*

Based on the limited data, it is difficult to make any assumptions about whether the fine structure of lightning on Uranus is similar to that on Saturn, what will be the characteristic durations of the structural components of lightning, and in which of them the maximum intensity of discharges will be concentrated. However, the dispersion can be determined quite accurately. The expected dispersion delay will be greater than it was in the Saturn observations, however this effect will be insignificant, because the interplanetary plasma between Saturn (at an average distance of 9.5 AU) and Uranus (19.2 AU) is on average several times less dense than between Earth and Saturn. This delay will not exceed a few hundred microseconds over a spectral range between 10 and a few tens of MHz.

To maximize the sensitivity to impulsive emission, the signal must be if possible integrated exactly over the emission duration and bandwidth. Shorter integration time increases the fluctuations $\sigma_{sky}$, whereas longer one dilutes the signal by averaging it with background noise only.

To detect flashes of a few to a few tens of millisecond duration, one may observe with 1-10 ms temporal resolution, neglecting the dispersion delay. Using a good spectral resolution (a few kHz) allows identification and elimination of man-made interference before integrating over the entire spectral range observed (tens of MHz). Flash detection thus requires moderate data volumes and simple processing, but one or several simultaneous OFF beams (see section 2.4.3) are necessary to distinguish signal from the source from local broadband interference (such as terrestrial lightning), and these measurements do not allow study of the fine temporal structure of the flashes (the bursts).

To detect and study this fine structure, measurements must be recorded with higher temporal resolution (e.g. 5-50 μs) and consequently coarser spectral resolution (200-20 kHz). Data can then be processed including a "blind" search by dispersion measure, as shown in Figure 3. As burst duration is likely larger than the temporal resolution of the observations and not much shorter than the dispersion delay, parametric de-dispersion followed by spectral integration over the entire bandwidth of the observations can be performed post-detection using a limited number of frequency channels in the dynamic spectra (500-1000), which greatly simplifies the processing.

Burst detection including parametric de-dispersion requires a processing heavier than flash detection, but several times more sensitive. A processing pipeline could combine both steps, first averaging in time high resolution data to search for flashes, then zooming at high resolution to study their fine structure.

A major improvement to the confidence than can be given to any impulsive signal detection, and thus to the sensitivity of the observations, is to observe simultaneously with two or more distant radio telescopes of similar sensitivities. Besides UTR-2, the low-frequency radio telescope NenuFAR (Zarka, et al., 2012b, 2015) is at an advanced stage of construction in Nançay (France) and is already 75% operational. Its compact core gathers 1824 dual-polarization antennas ensuring an effective area from 83000 $m^2$ at 15 MHz to 8500 $m^2$ at 80 MHz (https://nenufar.obs-nancay.fr/en/astronomer/).

The distance between UTR-2 and NenuFAR, over 2500 km, guarantees uncorrelated broadband interference environment (narrowband interference is easily removed by the data processing) and overlying terrestrial ionosphere. The above time-frequency resolutions of 5-50 μs and 200-20 kHz are easy to achieve both at NenuFAR (Zarka, et al., 2012b, 2015) and at UTR-2 (Zakharenko et al., 2016). The processed data can be compared in several ways: cross-correlation of time series within intervals of interest, comparison of the shape of broadband signals, etc. Comparison of the lightning signal parameters recorded with two different radio telescopes will thus provide more reliable criteria for the cosmic origin of the radiation than a threshold above the background noise in simultaneous ON and OFF beams from a single radio telescope.

Looking further ahead, Zarka et al. (2012a) generalized the criteria for detectability of planetary low-frequency radio signals from an ensemble of $N$ dipoles in space or on the Moon (preferably its far side, protected from Earth's interference). They showed that $N \sim 100$ is required for detecting SED, and $N \sim 1000$ for UED.

*4.4 Summary*

This paper has identified three scientific questions in ice giant atmospheric science, to be addressed by future missions and observations:
- Where in the ice giant atmospheres are the thunderstorms and what mechanisms charge them?
- What mechanisms cause the solar modulation of planetary brightness and where in the atmosphere do they act?
- What causes the differences between Uranus and Neptune in atmospheric electrical terms? Why does Uranus seem more electrically active?

A key theme has been to emphasise the significance of ground-based observations of the ice giants, both in terms of the solar modulation of their climate, and the possibility of lightning detection. Uranus lightning detection from Earth is possible in principle, and observations from any radio telescopes with suitable technical capability should be prioritised in the event of further storms.

Long-term observations are also important for these distant planets, as demonstrated in the telescope data sets of the planetary brightness discussed in section 3. As well as their slow seasonal variations, the ice giants, particularly Uranus, exhibit day to day variability and may also show annual or sub-annual cycles such as the ~30-year storm cycle on Saturn. Regular measurements over a long period of time are needed to capture the timescales of atmospheric variability. Unfortunately, the long-term brightness observations discussed in section 3 have recently ceased (Lockwood, 2019). Other long-term observations of ice giant meteorology are needed, which could be either ground or space-based (such as on the Moon, or on an orbiting telescope or interferometer).

Modelling of the cloud and aerosol microphysics in ice giant atmospheres is hindered by lack of data on the physical properties of cloud-forming materials in the relevant pressure and temperature range. Laboratory analogue experiments could help to explain the mechanisms behind thundercloud charging and ion-induced nucleation.

The ice giant systems are particularly fascinating worlds in atmospheric electrical terms, as both planets appear to have active lightning and solar-modulated climates. A simple lightning detector should be a scientific priority for an orbiter, and a combined electric field and conductivity sensor should form part of the atmospheric instrumentation carried by a descent probe. Ground-based observations and lab experiments can provide support and scientific progress to focus planning for the next mission.

**Acknowledgements**

TAN acknowledges support from the Jet Propulsion Laboratory, California Institute of Technology, under a contract with the National Aeronautics and Space Administration.

**Conflict of Interest**

The authors declare that they have no conflict of interest.

**Tables and Table Captions**

*Table 1*

|  | Distance from Sun [AU] | Number of detected whistlers | High frequency (HF) sferics | | |
|---|---|---|---|---|---|
|  |  |  | Detected events | Average flux at 1 AU [W m$^{-2}$ Hz$^{-1}$] | Source power [W/Hz] |
| Uranus | 19.2 | - | 140 | 6x10$^{-24}$ | 2.0 |
| Neptune | 30.1 | 16 | 4 | 1.4x10$^{-25}$ | 0.04 |

Table 1: Characteristics of Uranus and Neptune lightning detected by Voyager 2. The average flux and source power of the HF sferics represent the values around 15 MHz.

*Table 2*

| **Cloud material** | **Dielectric constant (at freezing point in K)** | **Source** |
|---|---|---|
| $CH_4$ | 1.7 (91) | Moses et al (1992) |
| $H_2S$ | 9 (187) | Gibbard et al (1999) |
| $NH_3$ | 25 (195) | Gibbard et al (1999) |
| $NH_4SH$ | ?? (261) | Gibbard et al (1999) |
| $H_2O$ | 80 (273) | Rinnert (1985) |

Table 2. Physical properties of cloud-forming materials in ice giant atmospheres.

*Table 3*

| Planet | Tropopause pressure (hPa), temperature (K) | Major atmospheric constituents (%) | Ion and electron concentration (cm$^{-3}$) | Ion (electron) mobility (m$_2$s$_{-1}$V$_{-1}$) | Positive (negative) conductivity (pS/m) |
|---|---|---|---|---|---|
| Neptune | 200, 55 | 80 H$_2$, 19 He | 10$_4$ | 2.2 x 10$_{-3}$ (1.8) | 4 (2870) |
| Uranus | 160, 55 | 83 H$_2$, 15 He | 10$_4$ | 2.8 x 10$_{-3}$ (2.4) | 5 (3820) |

Table 3. Estimated maximum atmospheric conductivity at the tropopause in a cloud-free atmosphere. Atmospheric parameters are from Mousis et al (2018) and charged particle concentrations from Capone et al (1977). Mobilities are estimated using data from Harrison and Tammet (2008) for ions and Pack and Phelps (1961) for electrons.

*Table 4*

| Planet | Description | 472 nm | 551 nm | Planet |
|---|---|---|---|---|
| **Uranus** | Best mechanism from physical modelling (Fraction of variance explained) | GCR (R$_2$=24%) | GCR (R$_2$=17%) | **Uranus** |
| | Pressure level at which τ=1; likely cloud type | 2000 hPa; H$_2$S ice cloud top | 3500 hPa; H$_2$S ice cloud | |
| **Neptune** | *As above* | UV (R$_2$=20%) | UV and GCR together (R$_2$=14%) | **Neptune** |
| | *As above* | 800 hPa; CH$_4$ ice cloud | 1000 hPa; CH$_4$ ice cloud bottom | |

Table 4. Summary of statistical and physical analysis of disk-averaged brightness fluctuations of the ice giants at two wavelengths (Lockwood and Jerzykiewicz, 2006), indicating the best estimate of the origin of the observed solar cycle variations. The top row shows the most likely mechanism for the solar cycle variation, and its coefficient of determination (R$_2$) (for Uranus, from Aplin and Harrison (2017) and for Neptune, from Aplin and Harrison (2016)). All statistical results quoted are significant to better than p<0.05. The second row indicates the pressure level at which the optical depth τ is unity (for Neptune, from Baines and Smith (1990) and for Uranus from Sromovsky et al (2011), and the likely cloud type at this level (Mousis et al, 2018).

**Figure Captions**

Figure 1: Uranian Electrostatic Discharges detected by the Voyager 2 PRA instrument. Panel a (bottom) shows a dynamic spectrum, panel b (right hand side) the number of UED as a function of frequency and panel c (top) the number of UED as a function of time (Reproduced with permission from Zarka and Pedersen, 1986).

Figure 2: Frequency-time spectrogram of a whistler recorded by the Voyager 2 plasma wave instrument at Neptune. The intensity is represented by the colour scale from blue (background intensity) to red (highest intensity). Reproduced with permission from Gurnett et al. (1990).

Figure 3: Data processing of radio signals from SED starting at Dec 23 2010, 03h56m27.0s UT. Top panel shows the dynamic spectra of SED with a time resolution of 7 $\mu$s. The middle panel shows the same data, after application of a post-detection de-dispersion procedure, expressed in terms of dispersion measure (DM) in parsecs cm$_{-3}$ and with the maximum (43 x10$_{-6}$ pc cm$_{-3}$) indicated as a horizontal line. The optimal DM was found by manually searching from DM = (10 to 100) x 10$_{-6}$ pc cm$_{-3}$ with a resolution of 10$_{-6}$ pc cm$_{-3}$. The bottom panel shows the Signal-to-Noise ratio (SNR) at the optimal de-dispersion.

Figure 4: Four times the galactic background fluctuation (4$\sigma_{sky}$) in Jansky (1 Jy = 10$_{-26}$ Wm$_{-2}$Hz$_{-1}$) as a function of receiver bandwidth (100 kHz to 10 MHz) and integration time (blue line for 20 ms, green line for 0.1 s). The average and the peak flux of Uranus lightning (UED according to Zarka and Pedersen, 1986) at Earth are indicated by a solid and a dashed black line, respectively.

Figure 5: The magnetic fields of Uranus and Neptune as measured by Voyager 2 (image provided by NASA)

Figure 6: Saturation ratio needed for condensation of $H_2S$ onto ions with between 1 and 10 elementary charges at temperatures corresponding to the Uranus cloud (a) top and (b) bottom with (c) showing the maximum number of charges that can be sustained on $H_2S$ droplets (solid line), with $H_2O$ at 273 K for comparison (dashed line).